\documentstyle[12pt]{article}
\begin{document}
\newcommand{\be}{\begin{equation}}
\newcommand{\ee}{\end{equation}}
\def\barr{\begin{array}}
\def\earr{\end{array}}
\newcommand{\ra}{\rightarrow}
\newcommand{\mr}{{\stackrel{<}{\sim}}}
\def\bi{\bibitem}
\def\lsim{\:\raisebox{-0.5ex}{$\stackrel{\textstyle<}{\sim}$}\:}
\def\gsim{\:\raisebox{-0.5ex}{$\stackrel{\textstyle>}{\sim}$}\:}
\def\gev{\; \rm  GeV}
\def\eg{ {\it e.g.}}
\def\ie{ {\it i.e.}}
\textwidth  15.5cm 
\title{
\begin{flushright}
IFT-96/14 \\ [1.5ex]
{\large \bf hep-ph/9608321 } \\ [1.5ex]
\end{flushright}
~\\
CONSTRAINING 2HDM BY PRESENT AND FUTURE $(g-2)_{\mu}$ DATA} 

\author{
      MARIA KRAWCZYK \\
{\it Institute of Theoretical Physics, University of Warsaw,
ul.Ho\.za 69} \\ 
{\it Warsaw, 00-681, Poland}
\and
JAN \.ZOCHOWSKI  \\
{\it Institute of Theoretical Physics, University of Warsaw,
ul.Ho\.za 69} \\
{\it Warsaw, 00-681, Poland}}
\maketitle
\newcommand{\tb}{\tan \beta}

\begin{abstract}

Constraints on the general  2HDM ("Model II") are obtained 
from the  
existing  $(g-2)_{\mu}$ data including limits on
Higgs bosons masses from LEP I data. 
We consider  separately  two cases: 
with a light scalar $h$ and with a light pseudoscalar $A$, 
assuming ${M_{h}+M_{A}} \ge {M_{Z}}$. 
The charged Higgs contribution is also included. 
It is found that already the present   $(g-2)_{\mu}$ data
 improve limits 
 obtained recently by ALEPH collaboration 
 on $\tb$ for the  mass of the pseudoscalar below $ \mr$ 2 GeV.
The improvement      
in the accuracy  by factor 20 in the  forthcoming 
E821 experiment  may lead to more stringent,
than provided by ALEPH group, limits up to $M_A\sim$ 30 GeV
if the mass difference between $h$ and $A$ is $\sim M_Z$.
Similar results should hold for a light scalar scenario
as well.

\end{abstract}

\section{Status of 2HDM.}

\subsection{Introduction.}
The mechanism of spontaneous symmetry breaking  proposed as
the source of mass for the gauge and fermion fields in the Standard 
Model (SM) leads to  a neutral scalar particle,
the minimal Higgs boson.  According to  the LEP I data,
based on the Bjorken process $e^+e^- \ra H Z^*$,
it   should be heavier than $66$ GeV\cite{hi},
also
the MSSM neutral Higgs particles have  been
constrained by LEP1 data to be heavier than 
  $\sim$ 45 GeV \cite{lep,susy,hi}. The general two Higgs doublet
model (2HDM) may yet accommodate a very light ($ \lsim 45 \gev$)
 neutral scalar $h$ {\underline {or}} a pseudoscalar $A$ as long as
$M_h+M_A \gsim M_Z$~\cite{lep}.

The  minimal extension of the Standard Model is to include
a second Higgs doublet to the symmetry breaking
mechanism. In two Higgs doublet models 
the observed Higgs sector is enlarged to five scalars: two
neutral Higgs scalars (with masses $M_H$ and $M_h$ for heavier and
lighter particle, respectively), one neutral pseudoscalar
($M_A$), and a pair of charged Higgses ($M_{H^+}$ and $M_{H^-}$). 
 The
neutral Higgs scalar couplings to quarks, charged leptons and gauge 
bosons are 
modified with respect to analogous couplings in SM by factors that 
depend on additional parameters : $\tan\beta$, which is
the ratio of the vacuum expectation values of the Higgs doublets
 $v_2/v_1$,
and the mixing angle in the neutral Higgs sector $\alpha$. Further,
 new couplings appear, e.g. $Zh (H) A$ and $ZH^+ H^-$.

In this paper we will focus on the  appealing version of the models
with two doublets ("Model II") where one Higgs doublet
with vacuum expectation value $v_2$ couples only to the "up"
components 
of fermion doublets while the other one couples to the "down" 
components \cite{hunter}. 
{{In particular,  fermions couple to the pseudoscalar $A$
with a strength  proportional to $(\tan \beta)^{\pm1}$
whereas the coupling of the fermions to the scalar $h$
goes as $\pm(\sin \alpha/\cos \beta)^{\pm1}$, where the sign
$\pm$  corresponds to  isospin $\mp$1/2 components}}. 
In such model FCNC processes are absent 
and  the $\rho $ parameter retains its SM value at the tree level.
Note that in such scenario 
the large ratio $v_2/v_1 \sim m_{top}/m_b\gg 1$ is naturally 
expected.
 
The well known supersymmetric model (MSSM) belongs to this  class.
In MSSM the relations among the parameters required by the 
supersymmetry appear, leaving only two parameters free
(at the tree level) e.g. $M_A$ and $\tb$.
In general case, which we call the general 2 Higgs Doublet Model
 (2HDM), masses and parameters $\alpha$ and $\beta$ 
are not constrained by the model.
Therefore the same experimental data may lead to very distinct 
consequences  depending on which 
 version of two Higgs doublet extension of SM,
supersymmetric or nonsupersymmetric, is considered.

\subsection  {Present constraints on 2HDM  from LEP I.}
Important constraints  on  the
parameters of two Higgs doublet extensions of SM were obtained
in the precision measurements at LEP I. 
The current mass limit on 
{\underline {charged}} Higgs boson $M_{H^{\pm}}$=
44 GeV/c was obtained at LEP I \cite{sob}
from process $Z \ra H^+H^-$, 
which is  { {independent}} 
on the  parameters $\alpha$ and $\beta$. 
(Note that in
the MSSM version one expect 
$M_{H^{\pm}} > M_W$). 
For {\underline {neutral}} Higgs particles $h$ and 
$A$ there are two
main and complementary sources  of information at LEP I. One  
 is the Bjorken processes $Z \ra Z^*h $
which constrains  $g_{hZZ}^2 \sim \sin^2(\alpha-\beta)$.
The second  process is   $Z\ra hA$,
constraining the $g_{ZhA}^2 \sim cos^2(\alpha-\beta)$ 
for $M_h+M_A\mr M_Z$
{\footnote {
 The off shell production could also be included,
 $\eg$ as in  \cite{susy}.}}.
 This Higgs pair production contribution depends 
also on the masses $M_h$, $M_A$ and $M_Z$.

Results on $\sin^2(\alpha-\beta)$ and $cos^2(\alpha-\beta)$   
can be translated into
the limits on neutral Higgs bosons masses $M_h$ and $M_A$.
 In the MSSM, due to relations among parameters, 
the above data allow to draw limits for the masses
 of {\underline {individual}}
 particles: $M_h\ge 45$ GeV for any $\tan \beta $ 
 and $M_A \ge$ 45 GeV for $\tan\beta \ge$1 \cite{susy,hi}.
In the general 2HDM the implications are quite different, here 
the large portion of the ($M_h$,$M_A$) plane,
where {\underline {both}} masses are in the range between 
0 and $\sim$50 GeV, is excluded \cite{lep}.

The third basic at LEP I process  in  search of a neutral  
Higgs particle is the Yukawa process, $\ie$
 the bremsstrahlung production   of the neutral Higgs
boson $h(A)$ from the heavy fermion, 
$e^+e^- \rightarrow f {\bar f} h(A)$, where $f$ means here
{\it b} quark or $\tau$
lepton.
This process plays a very important role since 
it  constrains the production of a very light pseudoscalar
even if the pair production is forbidden kinematically
{\footnote{neglecting the off shell production}},
 $\ie$ for $M_h+M_A>M_Z$.  
It allows also to look for a light scalar, being an additional
and in case of $\alpha=\beta$ the most important, source of information.
 The importance  of this process was stressed in many papers \cite{pok,gle},  
 the recent  discussion of the potential of the Yukawa process
 is presented in Ref.\cite{kk}.

{{ New analysis of the Yukawa process by  
 ALEPH collaboration \cite{alef} led to 
 the exclusion  plot (95\%)   on the $\tb$ versus the 
 pseudoscalar mass, $M_A$. (Analysis by L3 collaboration 
 is also in progress { \cite{l3prep}}.) 
 It happened that obtained  limits are rather weak
{\footnote{Note, that the obtained limits
are  much weaker than the limits estimated  in Ref.
\cite{kk}.}},
 allowing for the existence of a light $A$ with mass below 10 GeV
 with $\tb$ = 20--30 , for $M_A$=40 GeV $\tb$ till 100 is allowed
!
For mass above 10 GeV,
similar exclusion limits should in principle hold also for a 
scalar $h$ with 
the replacement $\tb\ra \sin\alpha/\cos\beta$.
Larger differences  one would expect however 
in region of lower mass,  where the production
rate for  the scalar is considerably larger   than 
for the pseudoscalar and therefore more 
stringent limits should be obtained \cite{kk}.

\subsection {The  2HDM  with a light Higgs particle.}

In light of the above results from precision experiments
at  LEP I
 there is still a possibility of the
 existence of one
light neutral Higgs particle with mass below $\sim$ 40--50 GeV.
As far as  other experimental data 
 are concerned, especially these from  low energy measurements,
they do  not  contradict 
this possibility as they 
cover only part of the parameter space of 2HDM, moreover
some of them  like the Wilczek process
 have large theoretical uncertainties
both due to  the QCD and relativistic corrections \cite{wil,hunter}
(see also discussion in \cite{bk,ames,mor}).

In following we will study the 2HDM assuming
that one  light Higgs particle  may exist.
Moreover we will  assume  according to  LEP I data
the following mass relation between the lightest
neutral Higgs particles: $M_h+M_A \ge M_Z$.
 We specify the model further  by choosing particular 
values for the parameters $\alpha$ and $\beta$
within the present limits from
LEP I. Since $\sin(\alpha-\beta)^2$ was found \cite{lep,hi}
to be
smaller than 0.1 for the $0\mr  M_h\mr$ 50 GeV,
and  even below 0.01 for a lighter scalar, 
we simply  take  $\alpha=\beta$.
This  assumption  leads to equal in  
 strengths of the coupling of  fermions to scalars   and pseudoscalars.
For the scenario with 
large $\tan\beta \sim {\cal O}(m_t/m_b)$  large
 enhancement 
in the coupling of both $h$ and $A$ bosons to the down-type
 quarks and charged leptons is expected.

As we described above the existing 
limits from LEP I for
a light  neutral Higgs scalar/pseudoscalar 
boson  in 2HDM are rather weak. 
Therefore it is extremely important to check if  more stringent
limits can be obtained from  other measurements.
The possible effect due to such a light Higgs
particle  at $ep$ collider HERA and at low energy LC,
as well  in heavy ion collisions at HERA and LHC are discussed elsewhere
\cite{bk,deb12,bol,war}.

In this paper we study in details the limits on the  2HDM from 
the precision $(g-2)$  experiment
for muon{\footnote {
Measurement of $(g-2)$ of the electron does not give
useful restriction due to small mass of the electron
 (see also \cite{haber}).}}.
We have studied this problem earlier using 
the simple approach 
constraining the individual
contributions $h$ (or $A$) 
(see
 Ref.\cite{ames,mor,deb12} and also \cite{gle}).
In the present analysis  we  take into account
the full contribution from 2HDM, i.e. exchanges of  
$h$, $A$ and $H^{\pm}$ bosons incorporating the present
constraints on Higgs bosons masses from LEP I.      
We study here the present data on $(g-2)_{\mu}$ \cite{pres} as well as 
the potential of the future E821 experiment \cite{fut}  
with the accuracy
expected to be more than 20 times better than the present one.

\section{Constraints on  the parameters of 2HDM 
from $(g-2)$ for the muon.}

\subsection{The room for new physics.}

The present experimental data limits on $(g-2)$ for muon 
averaged over the sign of the muon electric charge is given by \cite{data}:
$$a_{\mu}^{exp}\equiv{{(g-2)_{\mu}}\over{2}}=1~165 ~923~(8.4)\cdot 10^{-9}.$$
The quantity within parenthesis, $\sigma_{exp}$, refers to the uncertainty
in the last digit. The expected  high-precision E821 Brookhaven
experiment has design sensitivity of $\sigma_{exp}^{new}=
4\cdot 10^{-10}$ (later even 1--2 $\cdot 10^{-10}$\cite{czar})
 instead of the above $84\cdot 10^{-10}$.
It is of great importance to reach this accuracy in the theoretical
analysis.

The theoretical prediction of the Standard Model for this quantity 
consist of the QED, hadronic and EW
 contribution:
$$a_{\mu}^{SM}=a_{\mu}^{QED}+a_{\mu}^{had}+a_{\mu}^{EW}.$$
The recent  SM calculations of $a_{\mu}$
are based on the  QED results from \cite{qed,ki}, hadronic contribution
obtained in
\cite{mar,km,jeg,wort,ll,re,czar} and \cite{hayakawa} and
the EW results from \cite{czar,kuhto}. 
The uncertainties  of these
contributions differ among themselves considerably
(see \eg ~discussion below and in  Ref.\cite{nath,czar,jegwort}).
The main discrepancy 
is observed  for  the hadronic contribution,
therefore  we will consider here two cases : case A 
{\footnote {Refs.\cite{qed,ki,mar,km,ll,czar}}}
with relatively small error in the hadronic 
part and the case B {\footnote {Refs.\cite{qed,ki,jeg,hayakawa,czar}}}
 with the 2 times  
 larger error in the hadronic part.
 (We adopt here  the notation from \cite{nath}
 but one should be aware that the numbers used in
 this analysis differ slightly from Ref.\cite{nath}. 
 Basically our case A corresponds to case A from \cite{nath}, 
 whereas case B based on the \cite{czar}}.)

$$
\begin{array}{lrr}
case  &~{\rm {A~[in}}~ 10^{-9}]               &~{\rm {B~[in}}~ 10^{-9}] \\  
\hline
{\rm QED}   &~~~~~~~~1 ~165~847.06 ~(0.02)
  &~~~~~~~~~~~~~~~~~1 ~165~847.06 ~(0.02) \\
{\rm had}   & 69.70 ~(0.76) &  68.82 ~(1.54)  \\
{\rm EW}    & 1.51 ~(0.04)  &  1.51 ~(0.04)   \\
\hline
{\rm tot}   &1~165~9 1 8.27 ~(0.76)   & 1 ~165~9 1 7.39 ~(1.54)
\end{array} 
$$

\vspace{0.5cm}

Note that the hadronic contribution error dominates the
total error for the SM predictions and will influence strongly
the comparison with the new precision data from E821,
what will be discussed later.

The room for a new physics is given basically 
by  the difference between the experimental data and theoretical SM
prediction: $a_{\mu}^{exp}-a_{\mu}^{SM}\equiv \delta a_{\mu}$.
 In the following we will assume 
that it is consistent to use this difference $\delta a_{\mu}$
as the indication for the contribution due to Higgs particle(s)
of a beyond SM origin, like  2HDM,
although  in the calculation of
  $a_{\mu}^{EW}$ the (SM) Higgs scalar 
contribution is included.
This assumption based on the observation that
two EW calculations:
the EW result from Ref.\cite{czar}  based on $M_{Higgs}$=250 GeV
and the one   from Ref.\cite{kuhto} with  
$M_{Higgs}\ge$ 5 GeV,   differ only 
by $\sim 0.02 (0.06) \cdot 10^{-9}$.
{\footnote {The contribution due to the Higgs scalar for the  
$M_{Higgs}$=5 GeV was found to be of the level of 1 $\cdot 
10^{-11}$ \cite{kuhto}.}}
 
Below the difference $\delta a_{\mu}(\sigma) $ 
 for these two cases: A and B,
is presented together with 
 the error $\sigma$, 
 obtained by adding in quadrature the experimental 
and theoretical errors: \ie ~
$\sigma=\sqrt{(\sigma^2_{exp}+\sigma^2_{tot})}
\sim\sqrt{(\sigma^2_{exp}+\sigma^2_{had})} $. 

$$
\begin{array}{lcc}
case  &~{\rm {A ~[in}} ~10^{-9}]
	       &~~~~~~~~~~~~~~~{\rm {B ~[in}}~10^{-9}] \\  
\hline
\delta a_{\mu}(\sigma)    &4.73 (8.43) &~~~~~~~~~~~~~~5.61 (8.54)  \\
\hline
{\rm lim(95\%)} &-11.79\le\delta a_{\mu} \le 21.25 
     &~~~~~~~~~~~~~~-11.13\le\delta a_{\mu} \le 22.35\\
{\rm lim_{\pm}(95\%)} &-13.46\le\delta a_{\mu} \le 19.94&
~~~~~~~~~~~~~~ -13.71\le\delta a_{\mu} \le 20.84      
\end{array} 
$$

\vspace{0.5cm}

One can see that at one sigma level the difference $\delta a_{\mu}$
can be positive or negative.
For that  beyond SM scenarios
 in which both positive and negative 
$\delta a_{\mu}$ may appear,
 the 95\% confidence level (C.L.) bound can be calculated 
 straightforward (above denoted by $lim(95\%)$). 
For the model where  the contribution of
only {\underline {one}} sign 
is physically accessible,
 the other sign being unphysical, 
 ($\ie$ positive {\underline {or}} negative $\delta a_{\mu}$)
 the 95\%C.L. limits
should be calculated in different way \cite{data}.
These limits  calculated  separately for the positive and 
for the negative contributions 
are denoted above by $lim_{\pm}$(95\%).

We found that this latter approach leads to the sizeable
shift in the lower ($l_-$) and upper ($l_+$) 
limits 
with respect  to the standard (95\%) limits 
(by  -1.3 $\cdot 10^{-9}$ up to -2.6 $\cdot 10^{-9}$). 
That means that the possible negative contribution becomes larger
whereas the positive  becomes smaller when 
$lim_{\pm}$ method is used. 
The differences
between theoretical predictions (case A and B) for fixed method
of calculating confidence level
seem to be smaller that this effect.
All these effects may be important in future analysis.
\subsection{2HDM contribution to $(g-2)_{\mu}$.}

As we mentioned above the difference
 between experimental and theoretical value 
for the anomalous magnetic moment for muon
we ascribe to the 2HDM contribution, so in order 
to constrain the parameter space of the model
we take
 $\delta a_{\mu}= a_{\mu}^{(2HDM)}$.

To $  a_{\mu}^{(2HDM)}$ 
 a scalar $h$ ($a_{\mu}^h$), pseudoscalar $A$  ($a_{\mu}^A$)
 and the charged 
Higgs boson $H^{\pm}$ ($a_{\mu}^{\pm}$) contribute.
The relevant formulae can be found in the Appendix.
Each term  $a_{\mu}^{\Lambda}$ (${\Lambda}=h,~A ~{\rm or} ~H^{\pm}$)
disappears in the limit of large mass, 
at small mass the contribution reaches its maximum (or minimum if 
negative)
value.
The scalar contribution $a_{\mu}^h(M_h)$ is positive whereas the
pseudoscalar  boson $a_{\mu}^A(M_A)$ gives
negative contribution, also
the  charged Higgs boson contribution is 
negative. Note that  since the mass of $H^{\pm}$ is above 44 GeV 
(LEP I limit), 
its small contribution can show up
only if the sum of $h$ and $A$ contributions is small. 

We calculate the $  a_{\mu}^{(2HDM)}$
minimalizing the cross section in order to put limit
on the maximum allowed $\tb$.
Therefore  we take  $M_{\pm}^0$=44 GeV 
{\footnote {the case with mass equal 600 GeV  was also studied,
but in the range of masses of neutral Higgs bosons below 40 GeV
discussed here, the influence of so heavy  charged Higgs boson 
is negligible, see below.}}
and  
for mass of neutral Higgs bosons we assume  the relation
$M_h+M_A\ge M_Z$ as discussed before.\\ 

\noindent
We study 
two scenarios:                               
\begin{itemize}

\item {\sl a)} pseudoscalar $A$ is light,
 the scalar mass $M_h \ge M_Z-M_A$  and
we calculate the total 2HDM contribution as follows: 
$$   a_{\mu}^{(2HDM)}(M_A)= a_{\mu}^A(M_A)+ a_{\mu}^h(M_h^0=M_Z-M_A)
+ a_{\mu}^{\pm}(M_{{\pm}}^0)~~~~~~~(1a)$$
\item {\sl b)} scalar $h$ is light, the pseudoscalar mass is $M_A 
\ge M_Z-M_h$ and:

$$   a_{\mu}^{(2HDM)}(M_h)= a_{\mu}^h(M_h)+
 a_{\mu}^A(M_A^0=M_Z-M_h)+ a_{\mu}^{\pm}(M_{{\pm}}^0).~~~~~~~~(1b)$$

\end{itemize}

Due to opposite signs there appear a  cancellation in both scenarios
between scalar and pseudoscalar contributions, especially strong for
$M_A\sim M_h^0$ in (1a) or  $M_h\sim M_A^0$ in (1b).
Note that the total 2HDM contribution 
is for the  scenario {\sl a)}negative, whereas for the scenario {\sl b)}--
positive.
Therefore we have to include  this fact  calculating
 the 95\% C.L. bounds of $a_{\mu}^{(2HDM)}$.
This leads us to the limits $lim_{\pm}(95\%)$
described previously in Sec.2.1.

{\it Present data.}

Since the case A (Sec.2.1) gives  
more stringent $lim_{\pm}(95\%)$ constraints,
for 
{both} positive {\underline {and}} negative 
contributions, only this case will be used in the further analysis.
So, we have for the considered 2HDM scenarios 
allowed following ranges:
$${\sl a) } \hspace{4.5cm}-13.46 \cdot 10^{-9} \le 
  a_{\mu}^{(2HDM)} \le 0 ~~~~~~~~~~~~~~~~~~~~~~~~~~~~~~~~~(2a)$$
$${\it b) } \hspace{6.5cm}0\le   a_{\mu}^{(2HDM)} \le 19.94 
\cdot 10^{-9}.~~~~~~~~~~~~~~~~(2b)$$

Using Eqs.2a(2b)   the 95\% C.L. exclusion plots for $\tb$
versus $M_A$ ($M_h$) is calculated
 for  a light pseudoscalar (light scalar) scenario. 
 The results are presented in Fig.1 (2).
The regions above curves are  excluded.  

\begin{figure}[ht]
\vskip 5.45in\relax\noindent\hskip -0.8in
	     \relax{\includegraphics{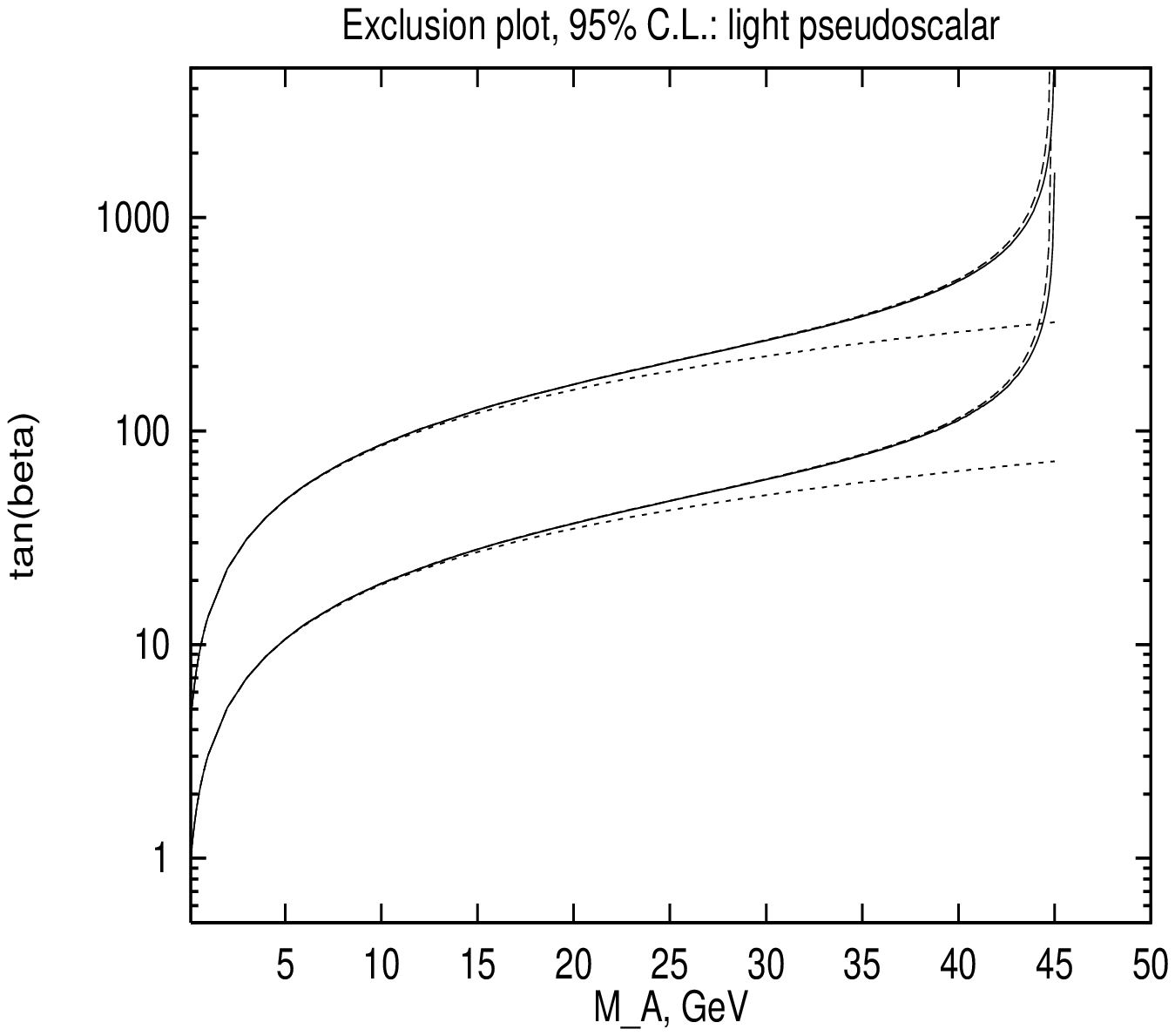}}
	     \relax\noindent\hskip 2.35in
	     \relax{\includegraphics{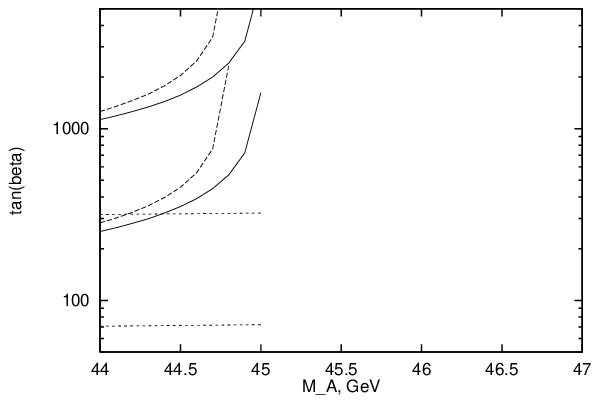}}
\vspace{-10.5ex}
\baselineskip 0.4cm
\caption{ {\em The  95\% exclusion plot, based on $lim_{\pm}$,
from the  $(g-2)_{\mu}$ data  
for pseudoscalar.
The present limits (Eq.2a) (upper curves) and for the improved  measurement
 (Eq.3a) (lower curves) are presented.
The total 2HDM (Eq.1a) contribution is represented by the solid lines,
 the short-dashed lines
correspond to the simple approach where only pseudoscalar contributes.
The long-dashed lines 
correspond to the case where the charged Higgs particle contribution
is neglected; for details see small figure, where the mass range
above 44 GeV is displayed.
} }
    \label{fig:fig1}
\end{figure}
\baselineskip 0.6 cm

The results based on the {\sl current}
data for $(g-2)_{\mu}$
are presented by {\sl upper} curves.  
The total contribution
of the 2HDM (solid line)
as well the simple approach 
where only
pseudoscalar (scalar) contribution (short-dashed line) 
is included  are presented in Fig.1 and 2.
They lead to similar limits up to mass $\sim$ 20 GeV.
Note that the simple approach for pseudoscalar (scalar) 
corresponds to the limits of negligible
contribution of scalar (pseudoscalar), i.e. to the large difference
in mass between $h$ and $A$ (above 150 GeV or so). 

\begin{figure}[ht]
\vskip 5.45in\relax\noindent\hskip -0.8in
	      \relax{\includegraphics{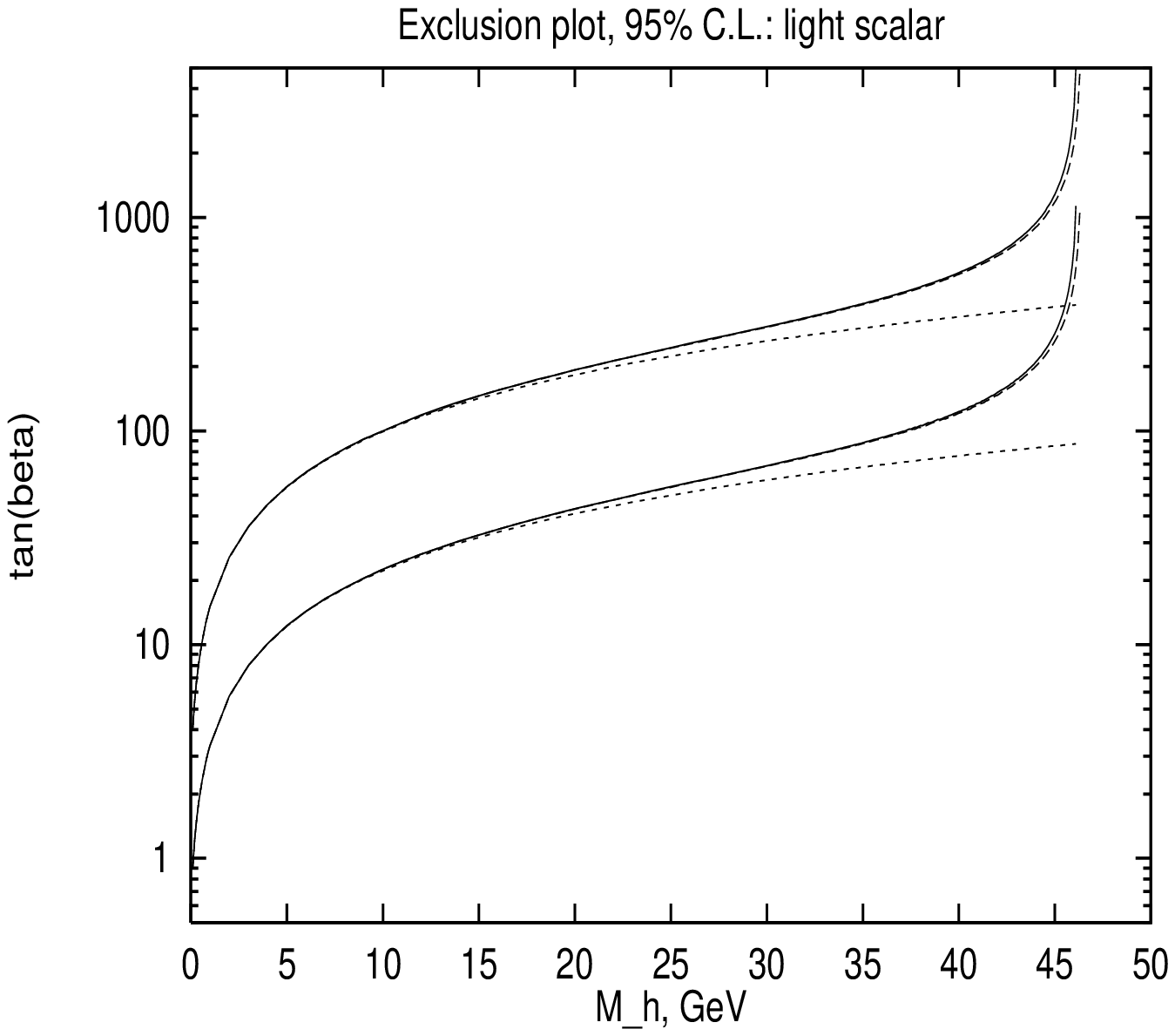}}
	     \relax\noindent\hskip 2.35in
	     \relax{\includegraphics{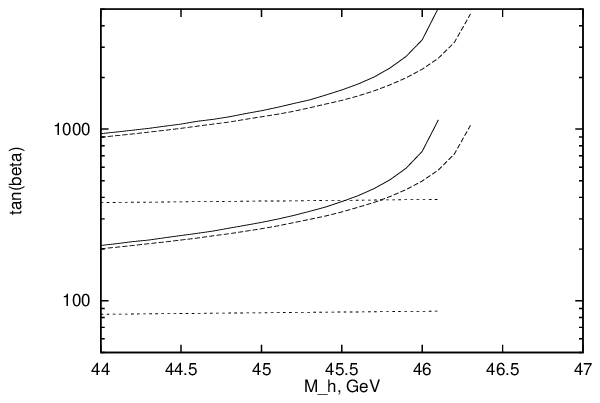}}
\vspace{-10.5ex}
\baselineskip 0.4cm
\caption{ {\em The  95\% exclusion plot, based on $lim_{\pm}$,
from the  $(g-2)_{\mu}$ data  
for  scalar.
The present limits (Eq.2b) (upper curves) and  the improved  measurement
 (Eq.3b) (lower curves) are presented.
The total 2HDM (Eq.1b) contribution is represented by the solid lines,
 the short-dashed lines
correspond to the simple approach where only scalar contributes.
The long-dashed lines 
correspond to the case where the charged Higgs particle contribution
is neglected; for details see small figure, where the mass range
above 44 GeV is displayed.
} }
    \label{fig:fig2}
\end{figure}
\baselineskip 0.6 cm

The charged Higgs boson contributes very little (in Figs.1 and 2
the difference between a solid line
and the long-dashed line, calculated without the charged Higgs 
boson term)
being visible only for masses of $h$ and $A$  above 40 GeV,
where the cancellation between scalar and pseudoscalar becomes
very strong (see  small figures in Figs.1, 2 for details). 

Interestingly,  the present $(g-2)_{\mu}$ data
can accommodate 
large value of $\tan\beta$ (20 or more)  for the Higgs boson masses 
equal or larger than 2 GeV
(see also the discussion in Ref.\cite{gle,ames,deb12}).

{\it Forthcoming data.}

Since  the dominate uncertainty in $\delta a_{\mu}$ (Sec.2.1)
is due to the experimental error,
the role of the forthcoming E821 experiment is  crucial
in testing the SM or  probing  a new physics.
We discuss now the potential of this measurement
for the constraining the 2HDM.

The future accuracy of the $(g-2)_{\mu}$ experiment 
is expected to be
 $\sigma^{new}_{exp}\sim0.4 \cdot 10^{-9}$ or better,
and one may in principle
calculate the expected error for the 
difference $\delta a_{\mu}^{new}$. Assuming no progress
in theoretical SM calculation  the above experimental accuracy
 will lead to
$\sigma^{new}$ = 0.86 and  1.6, for case A and B, respectively.

However, especially for the 
case A discussed in Sec.2.1, one could in principle 
expect such an improvement
in the calculation of the hadronic contribution
{\footnote {The improvement in the  ongoing experiments at low energy 
in expected as well.}} 
that the total uncertainty in $\delta a_{\mu}$ will be basically
due to the experimental error.
  We will take this point of view   
and in particular  
we will assume that 
the accessible range for $a_{\mu}^{(2HDM)}$, given presently by Eq.2a(2b)
 for a light pseudoscalar (scalar),
would be smaller by factor 20.
In this approach  the central value 
for the difference $\delta a_{\mu}$ is shifted by the same
factor 20 to lower value 0.24 $\cdot 10^{-9}$ (from 4.73 
$\cdot 10^{-9}$).
 So, we consider the following option (in $10^{-9}$):
$$
\delta a_{\mu}^{new} =  0.24, \hspace{0.5cm}{\rm and}\hspace{0.5cm} 
{\rm lim_{\pm}^{new}(95\%)}: -0.69\le\delta a_{\mu} \le 1.00  $$
or separately for two scenarios:
$${\sl a') } \hspace{4.5cm}-0.69 \cdot 10^{-9} \le 
  a_{\mu}^{(2HDM)} \le 0 ~~~~~~~~~~~~~~~~~~~~~~~~~~~~~~~~(3a)$$
$${\it b') } \hspace{6.5cm}0\le   a_{\mu}^{(2HDM)} \le 1.00 
\cdot 10^{-9}~~~~~~~~~~~~~~~~~(3b)$$
As expected much stringent limits on $\tb$ are obtained in this case
for pseudoscalar (scalar) 
-- see Fig.1 (2) {\sl lower} curves.

\subsection{Discussion and comparison with ALEPH limits.}

The comparison of the obtained 95\%C.L. exclusion plots 
based on the current and future accuracy for the 
$(g-2)_{\mu}$ data  
with the latest
results from ALEPH collaboration based on the 
Yukawa process \cite{alef}  is presented in Fig.3.
In this figure a simple approach prediction for 
a light $h$(solid line) or a light $A$ (long-dashed line) 
are shown for current and future $(g-2)_{\mu}$ 
measurement, {\sl upper} and {\sl lower} curves, respectively.

The results from ALEPH collaboration are presented by dotted line, 
only for the pseudoscalar case. 
 At very low mass, below say 2 GeV, the present $(g-2)_{\mu}$ data
 leads to stronger limits on $\tb$ than the Yukawa process.

 Similar results from Yukawa process 
as presented by dotted line
should also hold for the case of a light scalar  for $M_h$ above 10 GeV.
For lower mass both Yukawa and $(g-2)_{\mu}$ data may lead to comparable
exclusions plots.

Note that the predictions 
based on the simple approach 
can be justified if the mass difference between $h$ and $A$ is bigger
than, say 150 GeV.
In this case the improved $(g-2)_{\mu}$ data for muon will lead
 to more stringent limits over the whole considered range of masses.
If the mass difference between $h$ and $A$ is $\sim M_Z$
the improvement may be obtained  up to $\sim$ 30 GeV.

\begin{figure}[ht]
\vskip 5.5in \relax\noindent\hskip -0.9in
	      \relax{\includegraphics{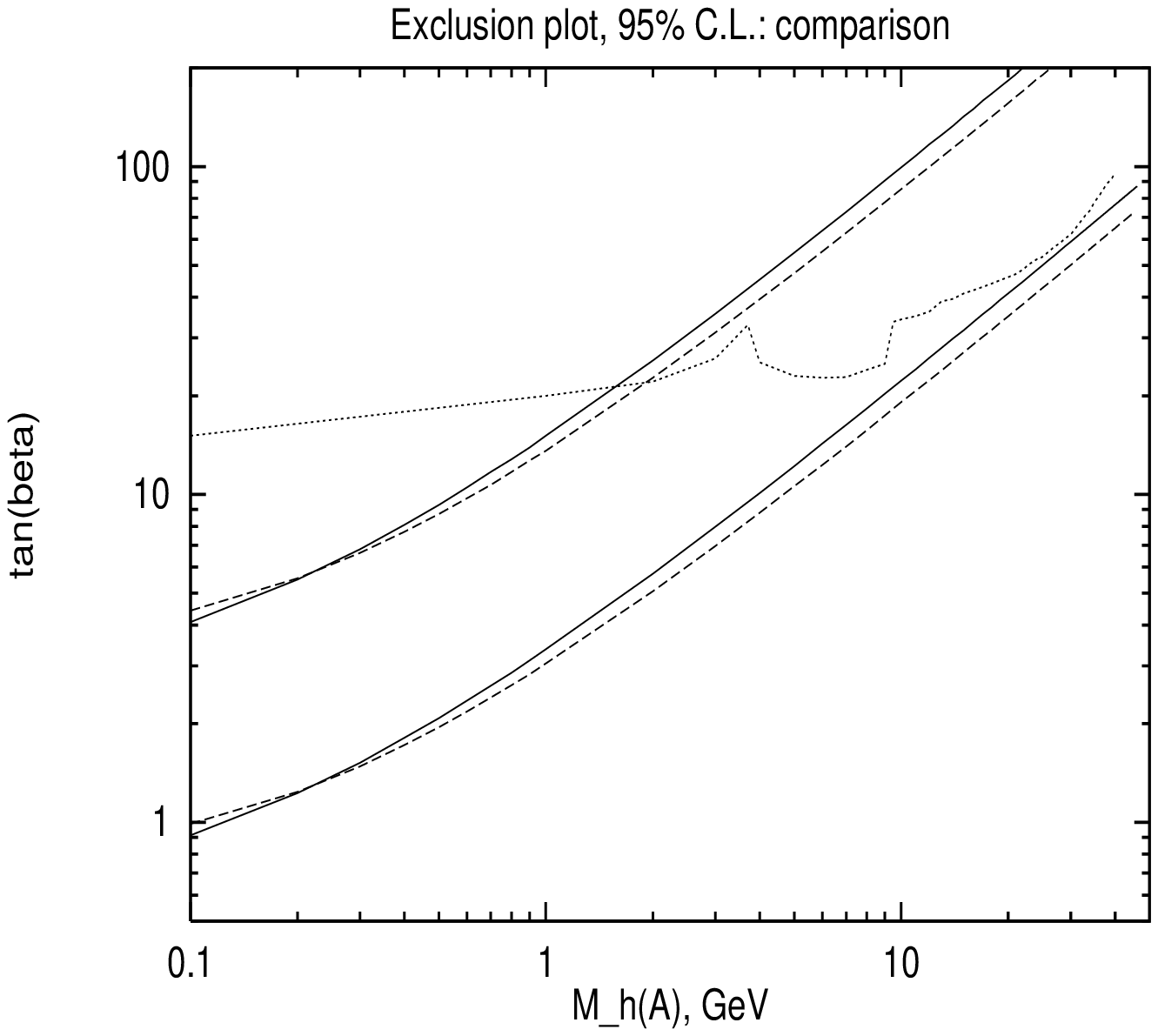}}
\vspace{-10.5ex}
\baselineskip 0.4cm
\caption{ {\em The  95\% exclusion plot, based on $lim_{\pm}$,
from the present (upper curves) and future 
(lower curves) $(g-2)_{\mu}$ data.
The simple approach
 results for a light $h$ (solid line),
and  a light  $A$ (long-dashed line)  
with  the new ALEPH
analysis for pseudoscalar (dotted line) are presented.
 } }
\label{fig:fig3}
\end{figure}
\baselineskip 0.6 cm

\section{Conclusion}
We studied the room for new physics 
in the   current muon anomalous magnetic moment 
data  based on
the newest theoretical calculations
of the SM prediction $a_{\mu}^{(SM)}$. 

The difference
 between experimental and theoretical value 
for the anomalous magnetic moment for muon
was ascribed to the 2HDM contribution, so
we took
 $\delta a_{\mu}= a_{\mu}^{(2HDM)}$.
From evaluated 95\% C.L. bounds for the positive or
negative value of 
 $a_{\mu}^{(2HDM)}$  we derived
constraints on the general  2HDM ("Model II").

We studied the total 2HDM contribution  $a_{\mu}^{(2HDM)}$
separately in  two cases: 
with a light scalar $h$ ($M_h\le$ 45) and with a light pseudoscalar $A$
($M_h\le$ 45), 
assuming according to the LEP I data that ${M_{h}+M_{A}}\ge {M_{Z}}$. 
A light scalar scenario leads to the positive, whereas the
one with a light pseudoscalar to the negative $a_{\mu}^{(2HDM)}$.
Large cancellation occure when both
mass of neutral Higgs bosons are nearly degenerated 
$M_h\sim M_A\sim M_Z/2$.
In this case 
the charged Higgs boson contribution may show up
 in the range of $M_h(A)$ above 40 GeV,
if we take the lower LEP I limit for $M_{\pm}$=44 GeV. 

We found that the contribution due to a light $A$ becomes larger
whereas this one due to a light $h$ becomes smaller by 
$\sim$ 1.5 or even 2.6 $\cdot 10^{-9}$
if the proper evaluation of 95\% C.L. is introduced.
 The differences
between theoretical predictions (mainly
due to the uncertainty in the hadronic corrections)
 on the level of 0.25 to 0.9 $\cdot 10^{-9}$
are smaller that this effect.
 Note also that the  the theoretical uncertainty
influences less the lower limit,
 relevant for the pseudoscalar contribution, 
than the upper one, where a scalar contribution is
expected.  
All these effects  may  be important for the 
future accuracy measurement of $(g-2)$ for muon.

It was found that already the present   $(g-2)_{\mu}$ data
 improve limits 
 obtained recently by ALEPH collaboration 
 on $\tb$ for low mass of the pseudoscalar:  $M_A \le$ 2 GeV.
The  
future improvement      
in the accuracy  by factor 20 in the  forthcoming 
E821 experiment  may lead to more stringent limits
than provided by ALEPH group up to $M_A$= 30 GeV
or higher for a larger mass difference between scalar and pseudoscalar.
Similar results should hold for a 2HDM with a light scalar.

\section{Acknowledgements}
We are grateful to Debayjoti Choudhury, Bohdan Grz\c{a}dkowski,
and Roman Nowak for useful discussions and comments.
We thank Jan Kalinowski for pointing us  Ref.\cite{jeg}.
Many thanks to 
J.-M. Grivaz, P. Janot and J.-B. de Vivie
for sending us the ALEPH results concerning the Yukawa process. 
Supported in part by the Polish Committee for Scientific Research.
\section{Appendix}
Theoretical predictions of an anomalous magnetic moment of muon 
from triangle vertex with the exchange of the Higgs 
boson ${\Lambda}$ (= $h$, $A$ or $H^{\pm}$) are given in
 Ref.\cite{lu,haber}.

\begin{displaymath}
{ a_{\mu}^{\Lambda}}=
{\frac{f_{\Lambda}^{2}}{8{\pi}^{2}}} {\tilde{L}_{\Lambda}},
\end{displaymath}
where the coupling constant ${f_{\Lambda}}$ is given by
\begin{displaymath}
{f}_{\Lambda}\equiv{{g ~m_{\mu}}\over{2~ M_{W}}} ~\tb,
\end{displaymath}
(in case of scalar $\tb$ should be replace by the $\sin(\alpha)/
\cos(\beta)$,
 but in our model we put $\alpha=\beta$, so the coupling
constant is universal ).

The integral ${\tilde{L}_{h(A)}}$ is 
for the neutral Higgs contribution given by:

\begin{displaymath} 
{\tilde{L}_{h(A)}}={\int_{0}^{1}}{dx}{\frac{Q(x)}{x^{2}+(1-x)
{({M_{h/A}}/{m_{\mu}})}^{2}}}. 
\end{displaymath}   
with

\begin{eqnarray}
{Q_{h}(x)}&=&{{x^{2}}{(2-x)}}\\
{Q_{A}(x)}&=&-x^{3}.
\end{eqnarray}

The charged Higgs particle exchange is described
by: 
\begin{displaymath}
{\tilde{L}_{\pm}}={\int_{0}^{1}}{dx}{-{x}(1-x)\over{x+
{({M_{\pm}}/{m_{\mu}})}^{2}-1}}. 
\end{displaymath}

\end{document}